\documentclass[12pt,preprint]{aastex}

\def\kms{\ifmmode{\rm km\thinspace s^{-1}}\else km\thinspace s$^{-1}$\fi}

\shortauthors{Torres et al.}
\shorttitle{OGLE-TR-33}

\begin{document}

\title{Testing blend scenarios for extrasolar transiting planet
candidates. I. --- OGLE-TR-33: a false positive}

\author{Guillermo Torres\altaffilmark{1},
	Maciej Konacki\altaffilmark{2},
	Dimitar D.\ Sasselov\altaffilmark{1},
	Saurabh Jha\altaffilmark{3}
}

\altaffiltext{1}{Harvard-Smithsonian Center for Astrophysics, 60
Garden St., Cambridge, MA 02138, USA}

\altaffiltext{2}{California Institute of Technology, Div.\ of
Geological \& Planetary Sciences 150-21, Pasadena, CA 91125, USA}

\altaffiltext{3}{Department of Astronomy, University of California,
Berkeley, CA 94720, USA}

\email{gtorres@cfa.harvard.edu}

\begin{abstract} 

We report high-resolution spectroscopic follow-up observations of the
faint transiting planet candidate OGLE-TR-33 ($V = 14.7$), located in
the direction of the Galactic center. Small changes in the radial
velocity of the star were detected that suggested initially the
presence of a large planet or brown dwarf in orbit. However, further
analysis revealed spectral line asymmetries that change in phase with
the 1.95-day period, casting doubt on those measurements. These
asymmetries make it more likely that the transit-like events in the
light curve are the result of contamination from the light of an
eclipsing binary along the same line of sight (referred to as a
``blend").  We performed detailed simulations in which we generated
synthetic light curves resulting from such blend scenarios and fitted
them to the measured light curve.  Guided by these fits and the
inferred properties of the stars, we uncovered a second set of lines
in our spectra that correspond to the primary of the eclipsing binary
and explain the asymmetries.  Using all the constraints from
spectroscopy we were then able to construct a model that satisfies all
the observations, and to characterize the three stars based on model
isochrones.  OGLE-TR-33 is fully consistent with being a hierarchical
triple system composed of a slightly evolved F6 star (the brighter
object) near the end of its main-sequence phase, and an eclipsing
binary with a K7-M0 star orbiting an F4 star. The application to
OGLE-TR-33 of the formalism developed to fit light curves of transit
candidates illustrates the power of such simulations for predicting
additional properties of the blend and for guiding further
observations that may serve to confirm that scenario, thereby ruling
out a planet. Tests such as this can be very important for validating
faint candidates. 
	
\end{abstract}


\section{Introduction}
\label{secintro}

Precise radial velocity searches around nearby, relatively bright
stars have yielded more than 100 extrasolar planet detections to date,
with orbital periods ranging from 2.5 days to nearly 15 years
\citep{Schneider:04}.  The discovery of transits in the case of
HD~209458b \citep{Henry:00, Charbonneau:00} made it possible for the
first time to measure the radius of the planet, as well as the
inclination angle that removes the spectroscopic $\sin i$ ambiguity in
the mass. This opened a new era in the study of these objects,
allowing astronomers to probe into the structure and evolution of
extrasolar giant planets, and leading even to the first detection of a
planetary atmosphere outside the solar system \citep{Charbonneau:02,
Vidal:03, Vidal:04}. 

With the realization that a fair number of the planets found by the
Doppler technique have relatively short orbital periods (a few days),
which makes the detection of transits across the disk of the parent
star much more likely, photometric searches for such events have
become a potentially important discovery tool. Numerous searches are
underway ---and others have already been completed--- that target a
variety of stellar populations including globular clusters, open
clusters, and the general field \citep[see][]{Horne:03}. The search
strategies of these programs are quite varied (field of view and
density of stars, magnitude limit, cadence of observations, passbands,
etc.), but they all aim at milli-magnitude photometric precision and
frequent observations to detect small drops in brightness of a few
percent with a typical duration of only a few hours. 

One of these searches, the Optical Gravitational Lensing Experiment
\citep[OGLE;][]{Udalski:92}, was not originally designed to search for
planets but in its latest incarnation (OGLE-III) it has turned out to be
one of the most successful in finding transit candidates. A total of
64 candidates have been reported by \cite{Udalski:02a, Udalski:02b,
Udalskietal:03} in three fields in the direction of the Galactic
center.  In many cases the candidates display light curves strongly
resembling that of HD~209458. A further 73 transit candidates were
reported in three other fields in the constellation of Carina
\citep{Udalski:02c, Udalskietal:03}. Dozens of other candidates have
been found by other teams, although only a fraction of them have
actually been reported in the literature. 

In general photometry alone cannot unambiguously establish whether the
object responsible for the transit is a planet (less than about 13
Jupiter masses), a brown dwarf ($\sim$13--80 Jupiter masses), or a
more massive stellar companion.  The problem is compounded by the fact
that the radius is nearly constant for objects with masses between
0.001~M$_{\sun}$ and 0.1~M$_{\sun}$ \citep[see,
e.g.,][]{Oppenheimer:00}.  Furthermore, light curves showing
transit-like events of similar depth as that produced, e.g., by a
Jupiter-size planet around a solar-type star ($\sim$1\%) can also be
produced by other astrophysical phenomena. These include grazing
geometry in a regular eclipsing binary, blending of the light of an
eclipsing binary with a single star that may or may not be physically
associated (and which dilutes the transit depth), or an eclipse
produced by a small stellar companion passing in front of a large
(early-type, or giant) primary star. Therefore, careful follow-up of
these candidates becomes crucial to establish their true nature,
especially given that the frequency of these ``false positive"
detections appears to be very high \citep[see, e.g.,][]{Latham:03,
Konacki:03a, Brown:03}. 

In this paper we focus on the transit candidate OGLE-TR-33, from the
first of the OGLE-III samples toward the bulge of the Galaxy. This
same sample led recently to the confirmation of the extrasolar
transiting planet around the faint ($V = 16.6$) star OGLE-TR-56
\citep{Konacki:03a, Torres:04}, only the second case known at the
time.  OGLE-TR-33 also appeared initially to be among the very best
candidates for harboring a planetary companion.  It has a photometric
period of 1.95 days, and a transit depth of 3.4\% in the $I$ band as
determined by the OGLE team.  However, as we describe below, follow-up
work shows this \emph{not} to be a transiting planet, but instead a
very interesting example of a blend configuration with an eclipsing
binary in a hierarchical triple system.  We use this case to
illustrate the importance of examining the spectral line profiles for
variable asymmetries, and we develop a technique to model the light
curve in detail in order to test various blend scenarios that satisfy
all observational constraints. Such models allow one to accurately
characterize the properties of the contaminating eclipsing binary, and
to predict other observable quantities. Techniques such as this are
valuable aids for validating transit candidates, particularly faint
ones, which are rapidly increasing in number as photometric surveys
continue operation. 

\section{Spectroscopic observations}
\label{sec:specobs}

OGLE-TR-33 ($17^{\rm h}56^{\rm m}41\fs19$, $-29\arcdeg 40\arcmin
05\farcs3$; J2000) was originally selected as a good candidate after
careful low-resolution spectroscopic reconnaissance of the entire
OGLE-III bulge sample carried out in 2002 \citep{Konacki:03b}. That
work was designed to reject cases with obvious stellar companions on
the basis of their large radial velocity variations (tens of~\kms) or
early spectral type.  Only 6 of the 39 candidates in the sample
subjected to this filtering survived, and were then observed at much
higher spectral resolution to try to detect low-amplitude velocity
variations presumably indicative of a companion of planetary mass.
The high-resolution observations of OGLE-TR-33 discussed in this paper
were conducted with the Keck~I telescope using the HIRES instrument
\citep{Vogt:94} on 2002 July 24--27. We obtained one spectrum on each
of 4 nights, with an exposure time of 30~minutes resulting in
signal-to-noise ratios of 15--20 per pixel. The wavelength coverage
(36 echelle orders) is approximately 385--620~nm at a resolving power
of $\lambda/\Delta\lambda \sim 65,\!000$.  The usual strategy for
Doppler planet searches with this instrument is to use the iodine gas
absorption cell as the wavelength fiducial, which makes it possible to
achieve m~s$^{-1}$ accuracy in the radial velocities for bright stars
\citep[e.g.,][]{Butler:03}.  Due to the faintness of our target ($V =
14.7$), however, the decision was made not to use the iodine cell in
this case in order to minimize light loss, essentially trading off
precision for increased signal.  For the wavelength calibration we
relied on exposures of a Thorium-Argon lamp taken before and after
each stellar exposure. The limit to the velocity accuracy attainable
with this procedure is ultimately set by the stability of the
spectrograph (see below).  The echelle reductions were performed using
the MAKEE package \citep{Barlow:02}, including scattered light
correction and cosmic ray removal. Orders blueward of about 400~nm
were not used due to low signal. 

\begin{figure}[!ht]
\vskip 1.5in
\includegraphics[angle=-90,scale=0.8]{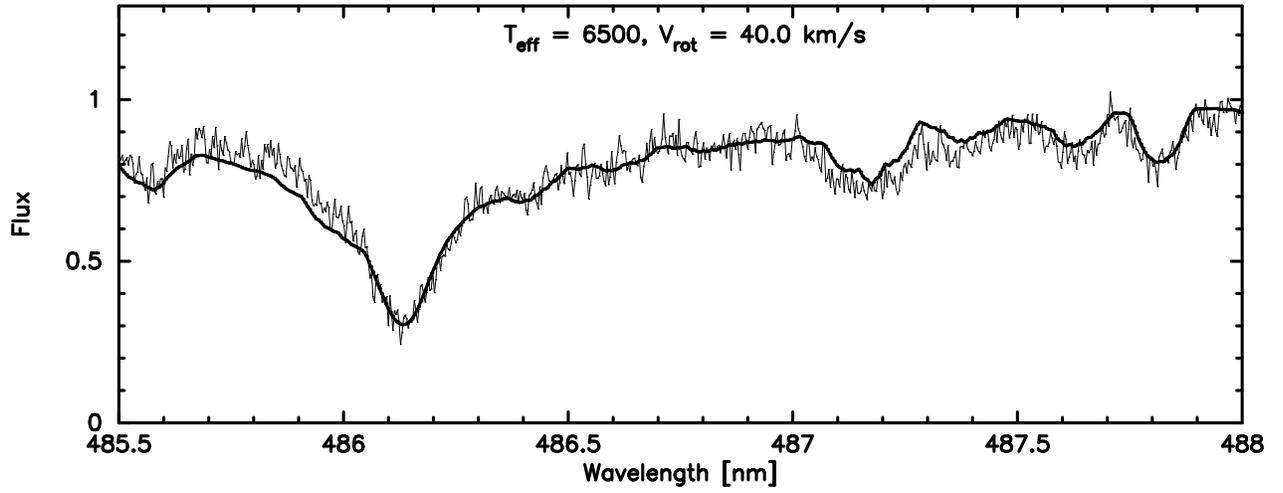}
\vskip 0.5in
 \figcaption[Torres.fig1.ps]{Observed spectrum of OGLE-TR-33 (obtained
by coadding the 4 individual spectra) compared to our best fit
theoretical spectrum (smooth line), used to derive the stellar
properties (see text).\label{fig:specfit}}
 \end{figure}

Radial velocities were obtained by cross-correlation using the XCSAO
task \citep{Kurtz:98} running under IRAF\footnote{IRAF is distributed
by the National Optical Astronomy Observatories, which is operated by
the Association of Universities for Research in Astronomy, Inc., under
contract with the National Science Foundation.}. The template used is
a synthetic spectrum chosen to closely match the star (which was
assumed to be of solar metallicity), and also broadened to match the
resolution of the instrument. The main parameters of this template
(effective temperature, $T_{\rm eff}$, and surface gravity, $\log g$)
were estimated by comparison with a high signal-to-noise spectrum made
by co-adding our 4 spectra. The model atmospheres are based on ATLAS9,
along with opacities and extensive line lists developed by
\cite{Kurucz:93}. We have added tools to compute the radial-tangential
macroturbulent velocity field broadening, as well as complex
rotational broadening of spectral lines.  The code has been found to
perform very well for solar-type stars (spectral type F--K).  We
fitted a large number of metal absorption lines of different
ionization and excitation stages, as well as the core and wings of the
Balmer H$\beta$ line. The parameters we derive correspond to a mid-F
star with $T_{\rm eff} \approx 6500$~K, and a rotational broadening $v
\sin i$ of about 40~\kms, quite typical of F stars in the field.  The
constraint on the surface gravity is weaker, but we estimate it is not
lower than $\log g = 3.5$.  Figure~\ref{fig:specfit} shows our fit to
the observed spectrum in the region of the H$\beta$ line. 

Our radial velocities for OGLE-TR-33 are the weighted average of the
31 orders we used, and have typical uncertainties of about 0.5~\kms,
except for the weaker exposure on the first night.  This is much
larger than we obtained for other OGLE targets during the same run
\citep[$\sim$100~m~s$^{-1}$;][]{Konacki:03b}, and is due to the large
rotational velocity of the star. 

Two stars with known planets were observed each night as velocity
standards: HD~209458 and HD~179949.  Because their spectroscopic
orbits are known \citep{Mazeh:00, Naef:04, Tinney:01}, we used them to
evaluate the stability of the instrument over the 4 nights of our run.
Suitable templates based on the same models as above were computed
using the stellar parameters reported for these stars in the
literature.  Velocity reductions of HD~209458 and HD~179949 with the
standard Th-Ar technique indicate that instrumental shifts are at the
level of 100~m~s$^{-1}$ or less
\citep[see][]{Konacki:03a,Konacki:03b}. 

\section{Photometric data}
\label{sec:photobs}

\begin{figure}[!ht]
\includegraphics[angle=0,scale=0.8]{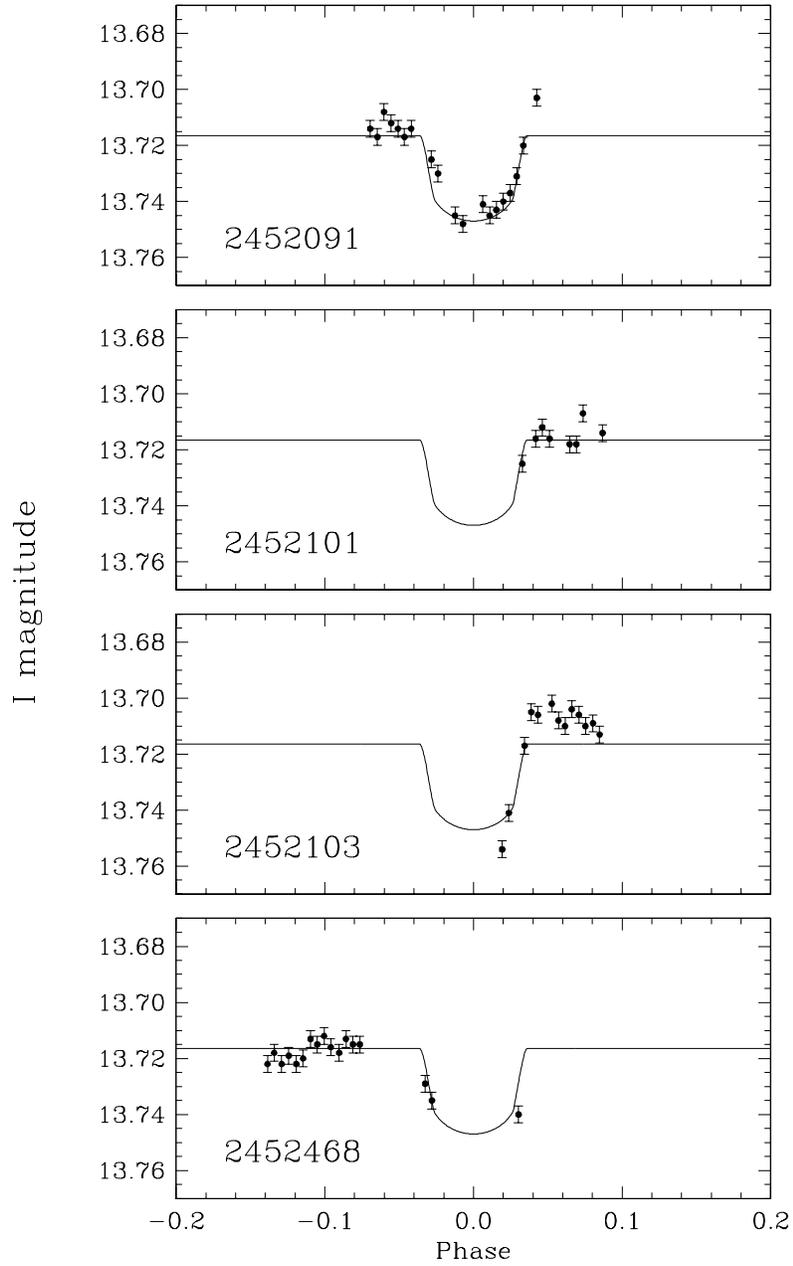}
\vskip 0.7in
 \figcaption[Torres.fig2.ps]{Individual transits recorded by the OGLE
team for OGLE-TR-33. Julian dates for the corresponding night are
indicate in each panel. The solid curve drawn for reference is a
transit model described in \S\ref{sec:specres} (see also
Figure~\ref{fig:transfit}).\label{fig:transits}}
 \end{figure}

OGLE-TR-33 was monitored photometrically by the OGLE group during the
2001 and 2002 seasons. A total of 396 observations were collected by
that team in the $I$ band\footnote{\tt
http://bulge.princeton.edu/$\sim$ogle/.}, with a nominal precision of
0.003~mag. In what follows, however, we have chosen to adopt a more
realistic value for the error of 0.006~mag, based on the scatter
outside of eclipse.  Four transit-like events have been recorded (see
Figure~\ref{fig:transits}), although only one is reasonably complete.
The transit ephemeris that has been derived from them\footnote{See
{\tt
http://bulge.princeton.edu/$\sim$ogle/ogle3/transits/transits.html}.},
and which we adopt in the following, is
 \begin{equation}
\label{eq:ephem}
T~{\rm (HJD)} = 2,\!452,\!062.49338 + 1.95320\cdot E,
\end{equation}
 where $E$ is the number of cycles elapsed since the reference epoch.
The depth of the transit as reported by the OGLE team is 3.4\%, and
the mean brightness of the system is $I = 13.71$, with an estimated
color index of $V\!-I = 0.95$ \citep{Udalski:02a}. 

The light curve of OGLE-TR-33 was examined independently by
\cite{Drake:03} and \cite{Sirko:03} for signs of variability outside
of eclipse (reflection effects, gravity brightening, ellipsoidal
variability), which would almost certainly indicate a stellar
companion as opposed to a planetary-mass object. Neither study found
any significant variations.  Further information on the nature of the
companion can be gleaned from the shape and duration of the transits
themselves, as shown by \cite{Seager:03}, which allow the properties
of the primary star to be inferred directly.  In principle this can be
used to rule out planetary companions in some cases, e.g., if the
stellar properties are inconsistent with those of normal stars or with
other information such as the spectral type. In practice the
usefulness of the test depends strongly on the precision and phase
coverage of the photometry, and unfortunately for OGLE-TR-33 the
ingress and the bottom of the transit are not very well covered.
While the resulting parameters for the star appear roughly consistent
with those of a main sequence object and would not seem to clearly
rule out a planetary companion, we consider these results to be
inconclusive in this case\footnote{\cite{Seager:03} have shown that
the stellar mass, radius, and density can be estimated readily by
careful measurement of the depth of the transit (relative change in
flux, $\Delta F$), its total duration ($t_T$), and the duration of the
flat portion ($t_F$), along with the knowledge of the orbital period.
For OGLE-TR-33 we measure those parameters to be $\Delta F = 0.034$,
$t_T = 0.074$, and $t_F = 0.050$ (the latter two in phase units),
although with large uncertainties due to incompleteness of the light
curve. The resulting stellar mass (1.74~M$_{\sun}$), radius
(1.56~R$_{\sun}$), and density ($\log [\rho/\rho_{\sun}] = -0.34$) are
roughly consistent with those of an F0 main sequence star
\citep[e.g.,][]{Cox:00}, allowing in principle for the possibility
that the companion is indeed a planet.  There may be a slight
inconsistency, however, between that spectral type and the effective
temperature for the star we estimated above (6500~K), and in addition
the inferred size of the companion is perhaps somewhat large for a
planet (0.29~R$_{\sun}$, or 2.8~R$_{\rm Jup}$). In view of the
difficulty in estimating the light curve parameters, and the fact that
the measured values are close to the limit of validity of the
equations of \cite{Seager:03}, we prefer to regard the above results
as inconclusive.}. 
	
\section{Spectroscopic results}
\label{sec:specres}

\begin{figure}
\epsscale{1.0}
\vskip -0.5in
\plotone{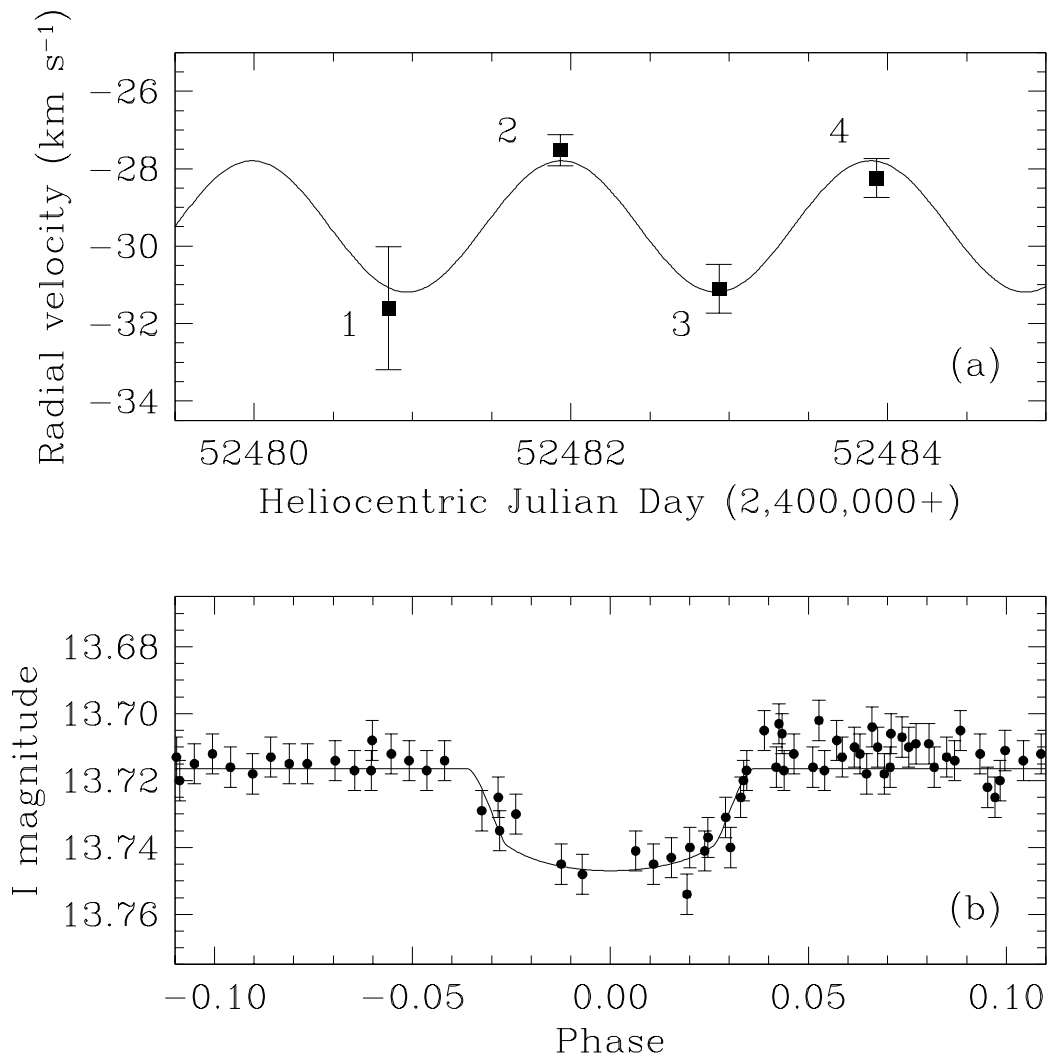}
\vskip -0.5in
 \figcaption[Torres.fig3.ps]{(a) Keplerian orbital fit to the formal
radial velocities of OGLE-TR-33. The phasing of the curve is fixed
from eq.(\ref{eq:ephem}), and only the velocity amplitude and an
offset have been adjusted. Observations are numbered to facilitate the
referencing with Figure~\ref{fig:bisectors}. (b) Transit light curve
of OGLE-TR-33, and our fit assuming the companion is a planet (see
text).\label{fig:transfit}}
 \end{figure}

Significant changes are evident in our Keck radial velocities for
OGLE-TR-33 over 4 consecutive nights, with a peak-to-peak variation of
nearly 4~\kms. As can be seen using eq.(\ref{eq:ephem}) the
observations were all taken near the quadratures, where the velocity
excursion is expected to be the largest.  In
Figure~\ref{fig:transfit}a we show those observations folded with the
known ephemeris. A Keplerian orbital fit is also shown, for an assumed
circular orbit as expected on theoretical grounds given the relatively
short period. Since the period and epoch are known sufficiently well
from the photometry, the only two remaining adjustable parameters are
the velocity amplitude, $K$, and the center-of-mass velocity,
$\gamma$. For these we obtained $K = 1.70 \pm 0.27$~\kms\ and $\gamma
= -29.49 \pm 0.27$~\kms. The amplitude is highly significant compared
to its error. For an edge-on configuration such as this would appear
to be, the formal mass we infer for the orbiting companion is $12.6
\pm 1.7$~M$_{\rm Jup}$, implying it is a large planet or perhaps a
brown dwarf. 

With the parameters adopted for the primary star in
\S\ref{sec:specobs}, a typical radius for an F star of 1.4~R$_{\sun}$,
and an appropriate linear limb-darkening coefficient in the $I$ band
of $u_I = 0.51$ \citep{Claret:00}, we fit the transit light curve
using the formalism of \cite{Mandel:02}.  The solution is shown in
Figure~\ref{fig:transfit}b along with the OGLE photometry. The
inclination angle derived is essentially 90\arcdeg, and the inferred
radius of the companion is $2.1 \pm 0.1$~R$_{\rm Jup}$. 

However, further investigation revealed spectroscopic evidence that
casts serious doubt on the interpretation that the velocities we
measured are the result of orbital motion of a planet or a brown
dwarf. We noticed, for instance, that the velocities were quite
sensitive to some of the details of how they were computed with XCSAO.
Specifically, they changed systematically and significantly as a
function of the fraction of the cross-correlation peak considered in
computing its centroid\footnote{The standard procedure followed in
XCSAO and other similar programs for computing the radial velocity is
to fit a parabola to the top portion of the cross-correlation peak,
above a certain fraction {\tt pkfrac} of the maximum, and to use that
fit to compute the centroid. The parameter {\tt pkfrac} can be changed
by the user, but defaults to 0.5.}. The velocity variations
(semi-amplitude $K$) seemed to increase as we included portions of the
correlation peak farther out on the wings, which is a rather strong
indication that the correlation peak (and therefore the mean line
profile) is asymmetric.  Furthermore, the asymmetry appeared to change
sides in phase with the velocities. Such behavior is not expected from
simple Keplerian motion of the star in its orbit.  Similar disproving
evidence has recently been found in at least two other cases that
appeared initially to be substellar companions as well
\citep{Queloz:01,Santos:02}. 

\begin{figure}
\vskip -0.8in\hskip -0.6in \includegraphics[scale=0.9]{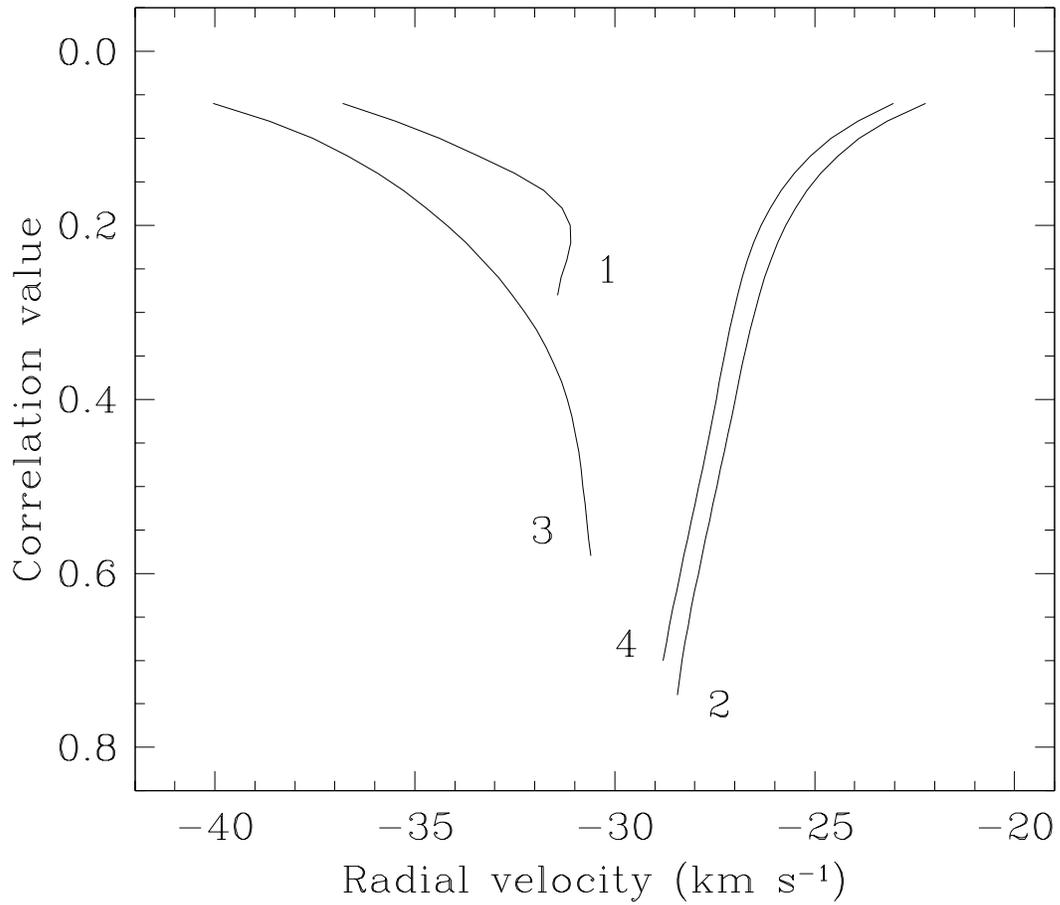}
 \vskip -0.4in \caption[Torres.fig4.ps]{Spectral line bisectors for
OGLE-TR-33 computed from the correlation function, which represents
the average spectral line. The core of the line is toward the bottom.
The profiles for each date are labeled as in
Figure~\ref{fig:transfit}a.  The asymmetries are obvious, and their
change in step with the phase is not normally expected from orbital
motion.\label{fig:bisectors}}
 \end{figure}

In order to examine the line asymmetries in OGLE-TR-33 more closely,
we computed the line bisectors \citep[see, e.g.,][]{Gray:92}
determined directly from the correlation function, which is
representative of the average shape of a spectral line for the star.
The individual correlation functions for each echelle order were added
together, resulting in an average line profile with much higher
signal. The line bisectors for each of our 4 nights are shown in
Figure~\ref{fig:bisectors} on an absolute velocity scale, and with the
core of the line toward the bottom.  In addition to the shifts, the
asymmetries (curved bisectors) are obvious and are quite large,
spanning several \kms.  For comparison, the line bisectors in the Sun
and similar stars (which display the well known ``C" shape due to
granulation) span only a few hundred m~s$^{-1}$.  More importantly,
the asymmetries in OGLE-TR-33 are seen to correlate with the phase in
the orbit (see Figure~\ref{fig:transfit}a), an effect that cannot
normally be produced solely by the orbital motion of the star. 

The most obvious explanation for the changing asymmetry is the
presence of another star in the system, whose spectral lines shift
back and forth with the photometric period, and distort those of the
primary.  Note that additional information is given by the type of
asymmetry observed. In this case it is ``leading" asymmetry (broader
wing of the line profile toward more positive velocities when the
measured velocity is also positive compared to $\gamma$, and
conversely on the other side of the orbit), as opposed to ``trailing"
asymmetry \citep[see also][]{Sabbey:95}. The latter case would result,
for example, from blending of the F star with another star having a
constant velocity near $\gamma$. This can be ruled out in this case,
and therefore the contamination must be from an eclipsing binary,
composed perhaps of a late F or early G star (causing the line
asymmetries) and an M star.  The systemic velocity of this binary must
be quite similar to that of the F star.  The binary would be
responsible for the transit-like events in the light curve, with the
original depth of the eclipses being diluted by the light of the main
F star. We examine this hypothesis in more detail below. 

\section{Testing the blend model}
\label{sec:blend}

A fair number of the light curves for extrasolar transiting planet
candidates found in various surveys have been explained before as
being the result of grazing eclipses, transits by a small star in
front of a large star, or blends with an eclipsing binary.
Unfortunately these situations appear to be far more common than
anticipated, and more common also than bona-fide transiting planets, a
fact that is just now beginning to be fully appreciated and that is
likely to impact design strategies for future transit searches
\citep[see][]{Brown:03}, even from space.  Nevertheless, relatively
few blend cases have been documented in detail in the literature
\citep[see][]{Mallen-Ornelas:03, Yee:02, Latham:03, Charbonneau:04}.
In OGLE-TR-33 there is rather strong evidence that contamination may
be the case as well, but we wish to go beyond qualitative explanations
and test this hypothesis in a more quantitative way. Our motivation
for doing this is to develop general procedures that may serve to test
other cases where the evidence for a blend is more subtle, or not
obvious at all. Such candidates are bound to present themselves in
current or future searches, and are the most challenging to rule out.
The shape and duration of the observed transits provide important
constraints on possible configurations involving a contaminating
eclipsing binary.  Therefore we first investigate whether the
OGLE-TR-33 photometry is consistent with this explanation, by
simulating realistic binary light curves including dilution effects.
Next we examine our spectra further for direct evidence of lines of
another star, guided by the results of the simulations. 

\subsection{Simulating eclipsing binary light curves}
\label{sec:ebop}

Although the variety of possible eclipsing binaries with which the F
star in OGLE-TR-33 could be blended may seem very large, clues from
the light curve itself and also from the spectroscopy limit the range
considerably. For example, the very short duration of the transits
($\sim$0.07 in phase, or 3.3 hours) indicates that the eclipsing
binary is well detached, as opposed to the fairly common W~UMa or
Algol systems. In addition, there is no evidence of a secondary
eclipse, suggesting the companion is much cooler than the primary. The
large asymmetries in the lines of the F star suggest that the primary
of the eclipsing binary is shifting back and forth by tens of~\kms,
and this would indicate that the companion is a star rather than a
substellar object.  Therefore we consider here only detached systems
composed of main sequence stars.  In order to generate synthetic light
curves to be fitted to the real data we relied on the computer program
EBOP \citep{Popper:81}, based on the Nelson-Davis-Etzel model
\citep{Nelson:72, Etzel:81}, which is adequate for well detached
systems. This model accounts for the oblateness of the stars, as well
as the effects of reflection, limb darkening, and gravity brightening.
In the following we refer to the primary and secondary of the
eclipsing binary as star~1 and star~2, while the F star OGLE-TR-33
itself (which provides ``third light" and dilutes the eclipses of the
binary) is referred to as star~3.  The physical properties of the
stars that are needed to generate a light curve are selected from
model isochrones, and the two components of the binary are assumed to
be on the same isochrone\footnote{The reliability of stellar evolution
models and their usefulness for predicting the physical properties of
stars is well supported by the strong observational constraints
provided by eclipsing binaries, at least in the range from about
1~M$_{\sun}$ to 10~M$_{\sun}$. For low-mass stars of spectral type M,
however, all existing evolutionary models become less realistic to
some extent due to limitations in the equation of state, convection
prescriptions, missing opacities, and/or the boundary conditions
adopted. We discuss these limitations and their impact on the present
analysis below.\label{foot:radii}}.  Whether or not star~3 can be
assumed to conform to the same isochrone as the binary will depend on
whether the stellar configuration is a physical triple system. If not,
the age ---and even the chemical composition--- can be different, and
a separate isochrone is used (see below). 

For each binary component the selected isochrone provides the radius
($R$), $\log g$, $T_{\rm eff}$, and absolute magnitudes in the visual
band ($M_V$) and in the $I$ band ($M_I$). We used these to compute
some of the parameters needed to generate a light curve with EBOP,
which are: $J$ (the central surface brightness of the secondary in
units of that of the primary), $r_1 \equiv R_1/a$ (the radius of the
primary in units of the semimajor axis, $a$), $k \equiv r_2/r_1$ (the
ratio of the radii), $q \equiv M_2/M_1$ (the mass ratio), and the
limb-darkening coefficients in the $I$ band (linear approximation).
The latter were interpolated from the tables by \cite{Claret:00} for
the $T_{\rm eff}$ and $\log g$ of each component.  The gravity
brightening coefficients were set to values appropriate for convective
stars (0.32). The orbit of the binary was assumed to be circular, and
the inclination angle in this particular case can be set to 90\arcdeg\
for simplicity and also because the flat-bottomed transits in
OGLE-TR-33 suggest total eclipses (although in principle the
inclination can be left as an additional free parameter; see below). 

The surface brightness ratio $J$ in the $I$ band was derived from the
expression
 \begin{equation}
J = {1\over k^2} {1-u_1/3\over 1-u_2/3} 10^{-0.4(M_{I,2}-M_{I,1})},
 \end{equation}
 where $u_1$ and $u_2$ are the limb-darkening coefficients for the
primary and secondary, and the last term represents the ratio of the
luminosities in the $I$ band in terms of the absolute magnitudes given
by the isochrones \citep[see, e.g.,][]{Kallrath:99}. 

Dilution of the depth of the binary eclipses by the light from star~3
is controlled by the ``third light" parameter, $l_3$, which is defined
in EBOP so that the sum of the three luminosities is unity.  Since the
absolute magnitude of star~3 is also given by the isochrones for its
adopted mass and age, third light (in the $I$ band) was computed
simply as
 \begin{equation}
\label{eq:thirdlight}
l_3 = \left[1+10^{-0.4(M_{I,1}-M_{I,3})}+10^{-0.4(M_{I,2}-M_{I,3})}\right]^{-1}.
 \end{equation}

For each trial combination of a primary and secondary for the
eclipsing binary, and the fixed properties of star~3 (see below), a
synthetic light curve can be generated and compared with the
observations in a least-squares sense. In practice we explore a wide
range of primary and secondary masses and search for a minimum of the
$\chi^2$ sum. 

The blend model described above is quite general and allows us to
predict a number of properties of the eclipsing binary components that
are potentially testable with the observations in hand. Given that the
primary is the most likely of the two stars in the eclipsing binary to
be detectable spectroscopically, one of the quantities of interest is
the velocity amplitude in its orbit, in \kms,
 \begin{equation}
\label{eq:kamp}
K_1 = 212.91\times P^{-1/3} {M_2\over(M_1+M_2)^{2/3}}~,
 \end{equation} 
 which follows directly from the masses indicated by the
isochrone (in units of the solar mass) and the known period $P$ in
days. 

Under the assumption that star~1 is rotating synchronously with the
orbital motion \citep[a reasonable assumption given that the period is
short; see][]{Hut:81}, the projected rotational velocity in terms of
the radius from the isochrone (expressed in units of R$_{\sun}$) is
given, in \kms, by
 \begin{equation}
\label{eq:vsini}
 v_1 \sin i = {50.61\over P} R_1 \sin i~.
 \end{equation}
 The relevance of this quantity here is that it can affect the
broadening of the spectral lines of star~1, and can make it more
difficult to detect, particularly if it is faint.  Although strictly
speaking the inclination angle in eq.(\ref{eq:vsini}) is that of the
spin axis, it can be safely assumed to be the same as that of the
orbit, or close to 90\arcdeg\ for all practical purposes
\citep{Hut:81}. 

Since the brightness of each star can also be predicted from the
isochrone, the light ratio between star~1 and star~3 (which is in
principle directly observable) is simply
 \begin{equation}
\label{eq:lratio}
L_1/L_3 = 10^{-0.4(M_{V,1}-M_{V,3})}~.
 \end{equation}
 We have expressed it here in the visual band rather than in the $I$
band because the observational constraint we have on this ratio comes
from our optical spectra, which are best matched to the $V$ band. 

Given that a large fraction of all field stars are in multiple systems
\citep[e.g.,][]{Duquennoy:91, Tokovinin:97, Tokovinin:02}, it may be
expected that in many cases the blend configuration will correspond to
a physical triple system.  In other blend situations the eclipsing
binary could be in the background or foreground. For convenience in
such cases we parameterize the relative distance between the binary
and the third star in terms of the difference in the distance moduli,
$\Delta M$, so that $d_{\rm bin} = d\times 10^{0.2 \Delta M}$ (where
$d$ is the distance to star~3).  The different distances for the
binary and star~3 introduce the need to account for differential
extinction, which can have a significant effect. For a given
combination of three stars it will change the amount of third light,
$l_3$, and also the predicted light ratio $L_1/L_3$.  Differential
extinction (e.g., in magnitudes per kpc) is a strong function of the
line of sight, and is essentially an unknown quantity because of the
unknown distribution of absorbing material in the Galaxy.  A
representative value often mentioned in the literature is
1~mag~kpc$^{-1}$. However, the true value is likely to be very
different in any direction one may choose, since dust appears to be
highly patchy not only across the sky but also in the radial direction
(particularly toward the Galactic center, where the OGLE fields are
located).  Nevertheless, we may adopt the prescription that the total
extinction up to a certain distance $d$ (in pc) is given by $A_V = a_V
(d / 1000)$, where $a_V$ is the coefficient of differential extinction
in the visual band (in mag~kpc$^{-1}$).  We also assume for simplicity
that $a_V$ is the same for the eclipsing binary as for star~3.
Extinction therefore adds two new unknowns to the problem: the
distance $d$ to star~3, and the coefficient $a_V$.  The distance to
the binary follows from the value of $\Delta M$. 

In order to solve for the new unknowns we make use of the constraint
provided by the total apparent magnitude of OGLE-TR-33 (comprised of 3
stars) in the visual and also in the $I$ band: $I = 13.71$ and $V =
14.66$ (the latter is derived from the $V\!-I$ color in
\S\ref{sec:photobs}). We require that the brightness of the three
stars (decreased by distance from the observer and by extinction) add
up to the measured values in both passbands, thus accounting for
differential reddening.  The combined light in the visual band is
 \begin{equation}
V = V_3 - 2.5\log\left[1+10^{-0.4(V_1-V_3)}+10^{-0.4(V_2-V_3)}\right]~,
 \end{equation}
 where the apparent magnitudes of the three stars are
 \begin{mathletters}
\label{eq:apparentmags}
\begin{eqnarray}
V_1 &=& M_{V,1} - 5 + 5\log d + \Delta M + a_V (d/1000) 10^{0.2\Delta M} \\
V_2 &=& M_{V,2} - 5 + 5\log d + \Delta M + a_V (d/1000) 10^{0.2\Delta M} \\
V_3 &=& M_{V,3} - 5 + 5\log d + a_V (d/1000)
\end{eqnarray}
 \end{mathletters}
 Similar expressions hold for the total magnitude in the $I$ band,
with a differential extinction coefficient $a_I$. For $a_I$ we adopt
the relation given by \cite{Cardelli:89}, $a_I = 0.48~a_V$ (assuming
$R_V = 3.1$)\footnote{\cite{Udalski:03} has recently reported
evidence of anomalous extinction towards the Galactic center, which
would imply a somewhat different value for $R_V$ as well as for
$a_I/a_V$. We do not consider these refinements essential here, as
they do not significantly change the results.}.  To account for the
effect of extinction in the contribution from third light $l_3$ and in
the predicted light ratio $L_1/L_3$, the absolute magnitudes in
eq.(\ref{eq:thirdlight}) and eq.(\ref{eq:lratio}) are replaced with
the apparent magnitudes given in eqs.(\ref{eq:apparentmags}) and their
analogs for the $I$ band. 

To summarize, the parameters to be adjusted in the most general case
are the masses $M_1$, $M_2$, and $M_3$, the age of the isochrone(s),
$i$, $d$, $\Delta M$, and $a_v$. In practice the ages and some of the
other parameters may be poorly constrained, depending on the quality
of the light curve, and may need to be fixed. A value for $M_3$ can
usually be adopted based on the spectral type or estimated effective
temperature of the star. 

We note, finally, that once a satisfactory fit to the light curve is
achieved, the formalism above makes it straightforward to generate a
curve based on the same physical parameters but for a different
passband. While the photometric signature from true planetary transits
should be independent of wavelength \citep[except for minor effects
due to limb darkening; see][]{Tingley:04}, the photometric signature
of a blend is in many cases color-dependent (but not always; see
below), and the predicted depth as a function of wavelength can be
tested observationally. 

\subsection{Application to OGLE-TR-33}
\label{sec:blendog33}

In our initial simulations for OGLE-TR-33 we assumed the three stars
form a physical triple system. This is based on the similarity between
the center-of-mass velocity of the binary and of that of the main
star, which we infer from the pattern of the spectral line asymmetries
(\S\ref{sec:specres}).  An isochrone was selected so as to match our
estimates of the properties of star~3, which we assumed to be an
unevolved main-sequence star. On the basis of its effective
temperature we adopted a mass of 1.3~M$_{\sun}$ \citep{Cox:00}, solar
composition, and a representative age of 2~Gyr. The isochrone was
chosen from the series of models by \cite{Girardi:00}\footnote{Our
choice for this application was based on the wide range of ages
available, and especially the fact that the isochrones reach down to
very low masses of 0.15~M$_{\sun}$, although as indicated in
footnote~\ref{foot:radii} the predicted properties of such low mass
stars are not very realistic.  Of particular importance for our
purposes is the fact that the predicted radii for such stars have been
shown to be underestimated by as much as 10--20\% \citep{Torres:02,
Ribas:03}.  In order to account for these discrepancies we have chosen
to apply an ad-hoc correction to the theoretical radii for low-mass
stars by careful comparison of these models with accurate
determinations in the eclipsing systems CM~Dra, CU~Cnc, YY~Gem, and
V818~Tau \citep{Metcalfe:96, Torres:02, Ribas:03}. These corrections
are typically 1.1--1.2, depending on mass, and they have a modest
effect on the depth of the eclipses.}. Although we obtained a
satisfactory fit to the OGLE light curve for a binary consisting of a
mid to late F star with $M_1 = 1.25$~M$_{\sun}$ (quite similar to
star~3) eclipsed by a late M star ($M_2 = 0.24$~M$_{\sun}$), some of
the predicted quantities from eqs.(\ref{eq:kamp})--(\ref{eq:lratio})
conflicted with our spectroscopic observations. In particular, we
predicted $L_1/L_3 = 0.81$ in the visual band, implying that we should
see a second set of absorption lines in our Keck spectra nearly as
strong as those of the main star.  While blending and the natural
broadening of the lines of the main star ($v \sin i \approx 40~\kms$)
and of star~1 (expected to have a similar rotation) could make this
detection slightly more difficult, it would be rather unlikely to be
missed, and additionally we would expect to see a flatter-peaked
cross-correlation function than we obtain. The evidence described in
the next section shows that the spectroscopic light ratio is in fact
much smaller ($\lesssim 0.2$), making this scenario untenable. No
reasonable change in the age of the isochrone removed this
discrepancy. 

Relaxing the condition that the three stars are physically associated
can also yield good fits to the light curve, for example if the
eclipsing binary is allowed to be in the background so that it is
fainter. One such solution is obtained for a binary system about 1~kpc
behind star~3, consisting of a mid or late F star ($M_1 =
1.28$~M$_{\sun}$) eclipsed by an M1 star ($M_2 = 0.45$~M$_{\sun}$).
The predicted light ratio in this case gives a more acceptable value
of $L_1/L_3 = 0.2$.  However, the probability of a chance alignment
such as this with a binary that happens to have a systemic velocity
very close to that of star~3 (as implied by the changing asymmetries
noted in \S\ref{sec:specres}) seems rather small. 

\begin{figure}
\epsscale{1.0}
\vskip -0.5in
\plotone{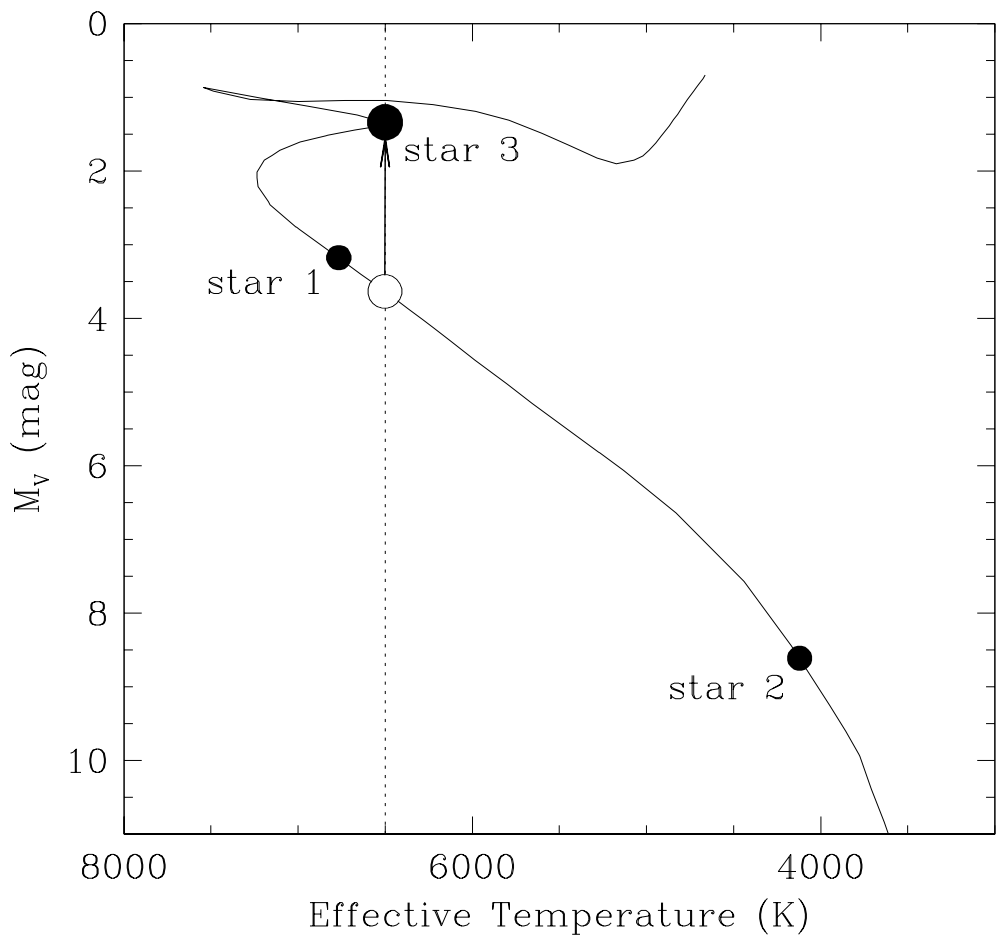}
\vskip -0.5in
 \figcaption[Torres.fig5.ps]{Isochrone from \cite{Girardi:00} for an
age of 1.12~Gyr and solar composition. The arrow shows that if star~3
is considered to be near the end of its hydrogen-burning phase (at the
start of the ``blue hook"; filled circle), it can be up to 2
magnitudes brighter than a star lower down the main sequence with the
same effective temperature (open circle). The dotted line represents
the 6500~K temperature we determine for OGLE-TR-33.  The locations of
star~1 and star~2 are the result of a blend model discussed in the
text.\label{fig:hr}}
 \end{figure}

An alternate way of making the eclipsing binary significantly fainter
than star~3, other than placing it in the background, is to make
star~3 itself intrinsically \emph{brighter}. The spectroscopic
constraints from our Keck spectra only allow us to state that
OGLE-TR-33 is not a giant star, but some degree of evolution is
allowed by the data as long as the temperature remains close to what
we estimate ($\sim$6500~K). In particular, if star~3 is beyond the
turnoff in the H-R diagram but still on the main sequence, near the
end of the hydrogen-burning phase, it can be considerably brighter
than an unevolved star with the same temperature.  This situation is
depicted in Figure~\ref{fig:hr}, where the arrow shows that a
difference in brightness of about 2 magnitudes is possible between an
evolved and an unevolved F star for an isochrone corresponding to an
age of $\log t = 9.05$ (1.12~Gyr), or close to it.  The evolved star
is of course considerably more massive in this case (1.97~M$_{\sun}$
instead of 1.3~M$_{\sun}$, as we had assumed before), but it still has
the same effective temperature. 

\begin{figure}
\epsscale{0.9}
\vskip -5.5in
\plotone{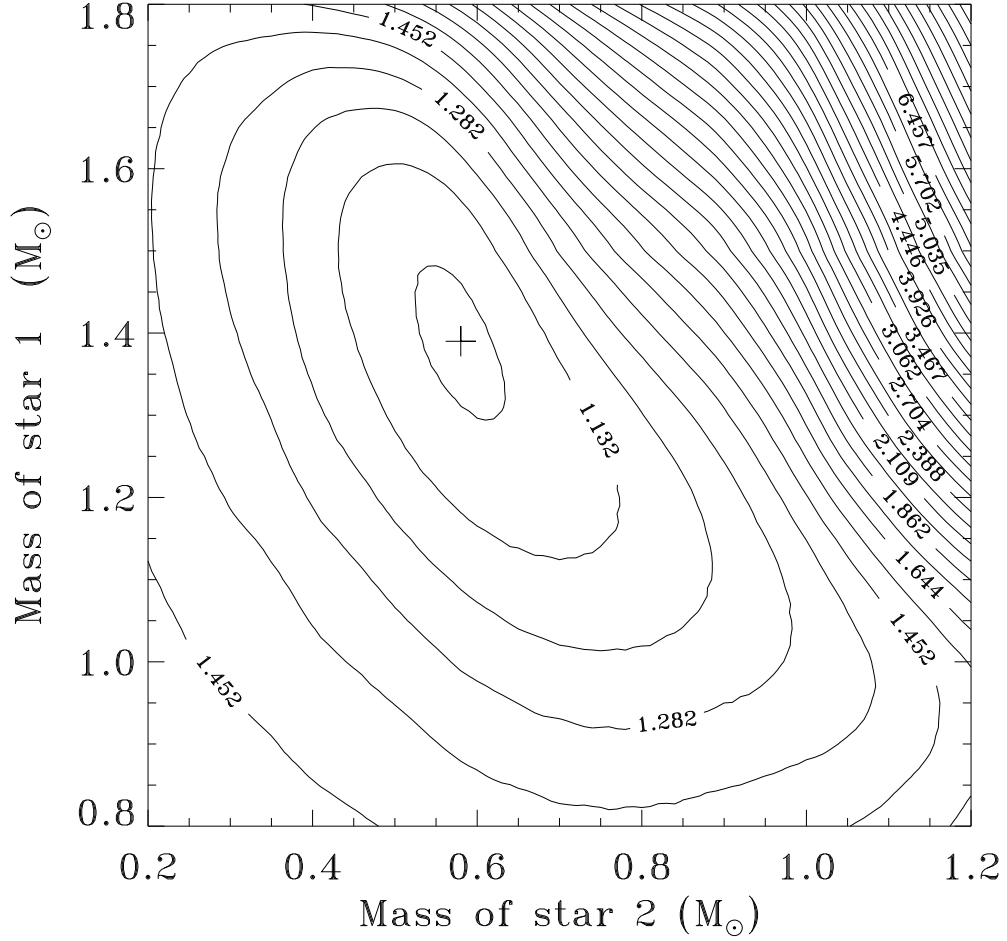}
 \figcaption[Torres.fig6.ps]{Contour plot of the $\chi^2$ surface
corresponding to a light-curve fit to the photometry for OGLE-TR-33
(star~3).  The model has the light of star~3 contaminated by an
eclipsing binary assumed to be physically associated (and therefore at
the same distance), composed of an F4 main sequence star and a K7-M0
star. The best fit masses are indicated by the plus sign (see text).
Properties for the three stars are constrained to follow an isochrone
with an age of $\log t = 9.05$ (1.12~Gyr), from \cite{Girardi:00}.
The location of the three stars in the H-R diagram is shown with
filled circles in Figure~\ref{fig:hr}.\label{fig:chi000}}
 \end{figure}

\begin{figure}
\epsscale{0.9}
\vskip -0.5in
\plotone{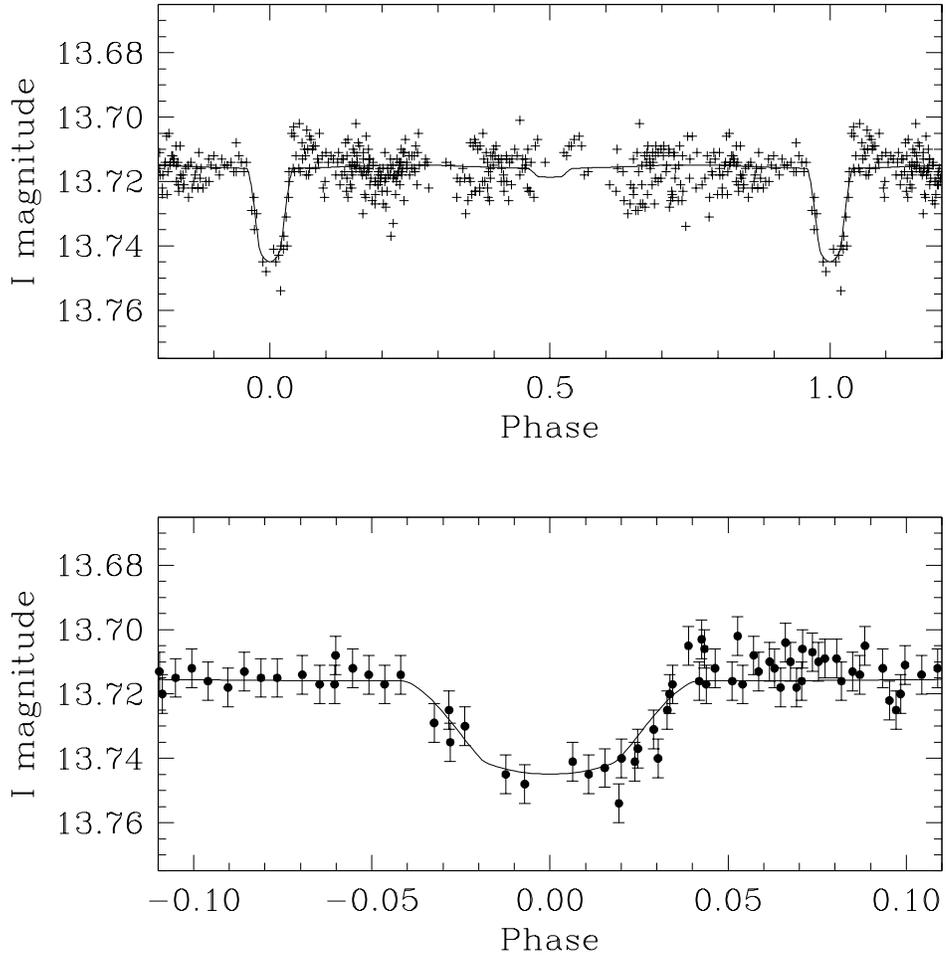}
\vskip -0.5in
 \figcaption[Torres.fig7.ps]{Photometry of OGLE-TR-33 together with
the light curve fit resulting from our blend model. In this model the
eclipsing binary and the main star are at the same distance (physical
triple system), but the main star is near the end of its main-sequence
life (see text).\label{fig:blendfit}}
 \end{figure}

We repeated the fit to the light curve for an evolved star~3 using
this isochrone, and we obtained a solution that is marginally better
than before but gives a predicted light ratio in the visual band
between star~1 and star~3 that is now much lower: $L_1/L_3 = 0.18$.
The $\chi^2$ surface as a function of the mass of star~1 and star~2
has a unique and well-defined minimum (see Figure~\ref{fig:chi000}),
and the fit to the photometry is shown in Figure~\ref{fig:blendfit}.
We note that given the quality of the data this solution is just as
satisfactory as the transit model fit in Figure~\ref{fig:transfit},
although the differences in shape are fairly obvious.  The blend model
predicts a very shallow secondary eclipse $\sim$0.003~mag deep, but it
is undetectable with the present photometry.  The resulting masses of
the stars comprising the eclipsing binary are $M_1 = 1.39$~M$_{\sun}$
\citep[spectral type approximately \ion{F4}{5};][]{Cox:00} and $M_2 =
0.58$~M$_{\sun}$ (\ion{K7-M0}{5}), and their locations on the
isochrone are shown in Figure~\ref{fig:hr}.  Star~1 is slightly hotter
than star~3, but fainter. The predicted velocity semi-amplitude and
projected rotational velocity of star~1 are $K_1 = 63$~\kms\ and $v_1
\sin i = 40$~\kms. The distance to the triple system is determined to
be 3.5~kpc, and the differential extinction coefficient that fits the
combined colors is $a_v = 0.23$~mag~kpc$^{-1}$. Tests with a range of
inclination angles indicated that the preferred value is in fact
90\arcdeg.  Other parameters of the fit are $r_1 = 0.186$, $k =
0.391$, and $J = 0.121$.  Star~3 contributes 85\% of the light in the
$I$ band. 

The blend model also predicts that the eclipse depth in the $V$ band
should be slightly larger than in $I$, but by less than 0.003~mag, a
difference that is probably too small to be detected with any
confidence in future observations unless the measurement precision is
significantly improved. The reason for this is the similarity between
the colors of star~1 and star~3, both of which are F stars. Thus, in
this particular case the color dependence of the eclipse depth would
\emph{not} be a useful diagnostic to rule out a planetary companion. 
		
\subsection{The lines of another star}
\label{sec:todcor}

The blend scenario in the preceding section has allowed us to make
quantitative predictions about the primary of the eclipsing binary
that can be tested with our data. We expect that if this star is
visible in our spectra, it will be roughly 5 times fainter than the
main star and its lines will move back and forth with semi-amplitude
$K_1$. Although our initial visual inspection of the spectra and
cross-correlation functions did not reveal any obvious signs of double
lines, in hindsight the asymmetries themselves are a clue to the
presence of star~1. Given that the lines of the main star are
significantly broadened by rotation ($v \sin i \approx 40$~\kms), and
that we expect the $v \sin i$ of star~1 to be about the same, the
spectral lines of the two stars are likely to be heavily blended at
all phases since the semi-amplitude $K_1$ is of the same order as the
combined rotational broadening. In cases such as this traditional
spectral analysis techniques are not very useful. 

We re-examined our spectra using TODCOR \citep{Zucker:94}, a
two-dimensional cross-correlation technique that has the ability to
use a different template for each star in a composite spectrum.  With
TODCOR it is often possible to measure accurate velocities for both
stars even when the lines are severely blended.  For the main star we
used the same synthetic template as before (\S\ref{sec:specobs}), and
for star~1 we generated a synthetic spectrum based on the parameters
predicted by the blend model: $T_{\rm eff} = 6750$~K, $\log g = 4.0$,
$v \sin i = 40$~\kms, and an assumed solar metallicity. The
correlation functions for the individual echelle orders were combined
together in a manner analogous to the technique introduced by
\cite{Zucker:03}. Due to the similarity in temperature between the two
stars (corresponding approximately to spectral types F4 and F6) the
light ratio was assumed to be constant with wavelength\footnote{Checks
with low-resolution spectra of the appropriate temperatures indicate a
change of only 10\% in the flux ratio across the entire spectral
window, a difference that is unimportant for our purposes.}. 

\begin{figure}
\epsscale{0.9}
\vskip -0.1in
\plotone{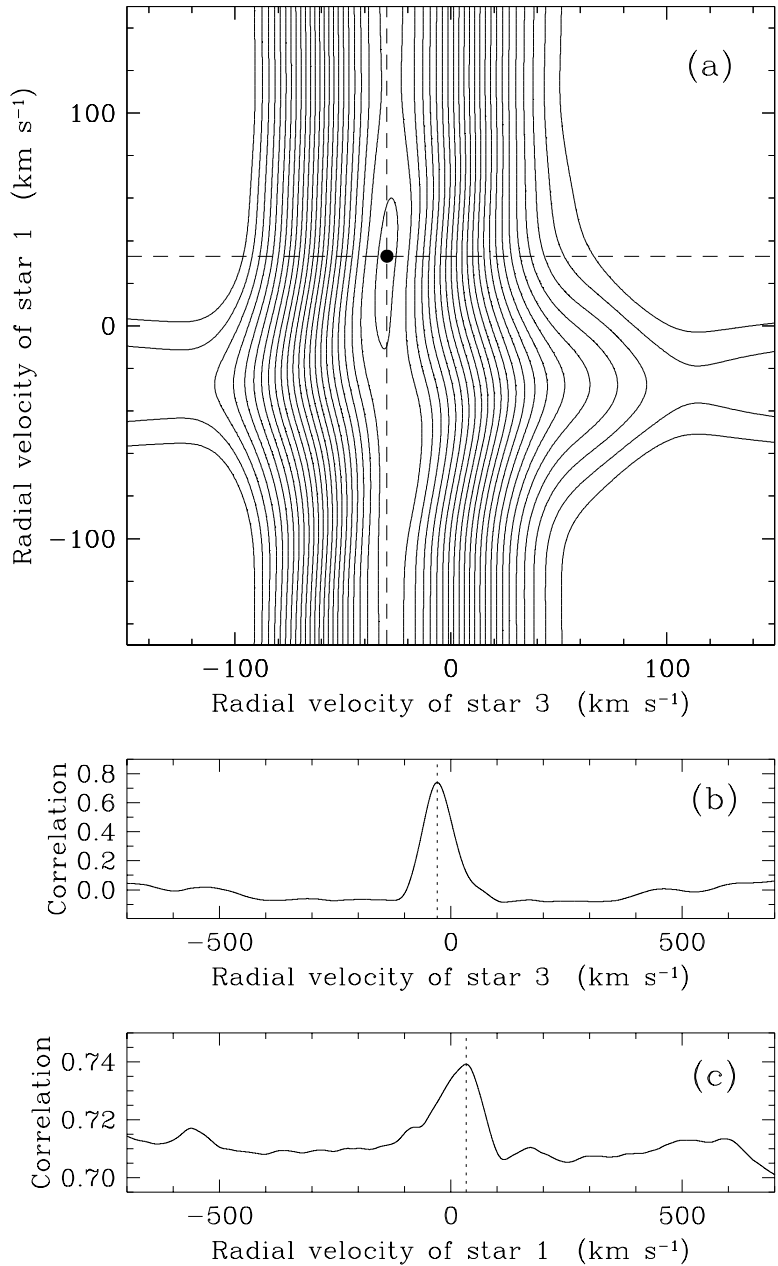}
\vskip 0.8in
 \figcaption[Torres.fig8.ps]{(a) Contour diagram of the
two-dimensional cross-correlation surface computed with TODCOR for the
last of our observations, as a function of the velocities of star~1
and star~3.  The maximum is indicated by the dot, and the dashed
lines represent cuts through the surface that are shown in the lower
panels.  (b) Cross section of the two-dimensional correlation function
at a fixed velocity for star~1, showing the peak corresponding to the
main star in OGLE-TR-33. The measured velocity is indicated by the
dotted line. (c) Same as (b) at a fixed velocity for star~3, showing
the clear detection of star~1. \label{fig:todcor}}
 \end{figure}

\begin{figure}
\epsscale{0.9}
\vskip -0.5in
\plotone{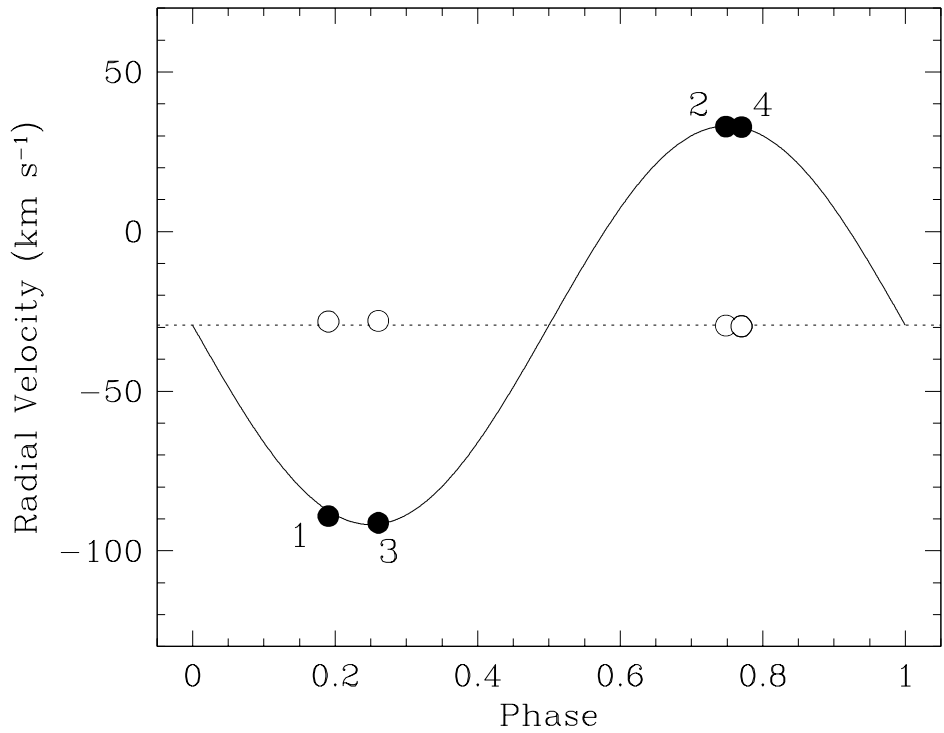}
\vskip -0.5in
 \figcaption[Torres.fig9.ps]{Radial velocities of star~1 (primary of
the eclipsing binary blended with OGLE-TR-33) measured with TODCOR
(filled circles).  Dates are labeled as in Figure~\ref{fig:transfit}a.
Also shown is the corresponding spectroscopic orbital solution
adopting the ephemeris from eq.(\ref{eq:ephem}), with the dotted line
representing the center-of-mass velocity, $\gamma$. The velocities for
star~3 are shown with open circles, and lie very close to $\gamma$.
The error bars are smaller than the symbols.\label{fig:star1orbit}}
 \end{figure}

As illustrated in Figure~\ref{fig:todcor} the lines of star~1 are
indeed clearly present in our spectra. The top panel shows a contour
plot of the two-dimensional cross-correlation surface, as a function
of the velocities of star~1 and star~3. The maximum correlation is
indicated by the dot, and lies along a steep ridge line that runs
vertically through the diagram, which is produced by the large
difference in brightness between the two stars. Cuts through this
surface at fixed values for the velocities of each star are indicated
by the dashed lines, and are shown in the lower two panels. The
obvious peak in the lower panel is evidence that star~1 is clearly
detected, and we were able to measure its velocity on all four of our
spectra. The radial velocities we derive for both stars are given in
Table~\ref{tab:rvs}.  A spectroscopic orbital solution for star~1 from
these velocities assuming a circular orbit and a fixed ephemeris from
eq.(\ref{eq:ephem}) gives a semi-amplitude of 62~\kms, which is very
close to the predicted value of 63~\kms. This orbit is shown in
Figure~\ref{fig:star1orbit}.  The spectroscopic light ratio
(approximately in the visual band) measured with TODCOR is 0.15, also
reasonably close to the prediction. The velocities for star~3, on the
other hand, show much less variation than our original one-dimensional
results from \S\ref{sec:specres} (see also Figure~\ref{fig:transfit}).
The {\it rms\/} residual from the mean is now 0.87~\kms, compared to
2.03~\kms\ for the velocities shown in Figure~\ref{fig:transfit}. This
is an indication that the use of TODCOR has effectively removed the
distortions (asymmetries) that initially suggested a velocity
variation. We note also that the weighted average of the new
velocities for star~3, $-28.95~\kms$, is very close to the
center-of-mass velocity of the orbit for star~1, which is
$-29.34~\kms$. This is just as expected for a physical triple system. 

Given the uncertainties and assumptions in the blend model, the
agreement in the properties of star~1 and the overall consistency of
the results is quite remarkable, and it is entirely possible that an
even closer agreement could be reached by fine-tuning various
parameters of the model, although we have not considered this
necessary here. 

\section{Blending across the H-R diagram: a cautionary tale}

The example of OGLE-TR-33 presents a warning about the kinds of blend
scenarios that may be the most challenging to expose.  As we showed in
\S\ref{sec:blendog33} the slightly evolved nature of the main star
allows it to be significantly brighter intrinsically, and this in turn
makes it easier to hide a fainter eclipsing binary in its glare that
would typically be difficult to detect spectroscopically. 

\begin{figure}
\epsscale{0.9}
\vskip -0.5in
\plotone{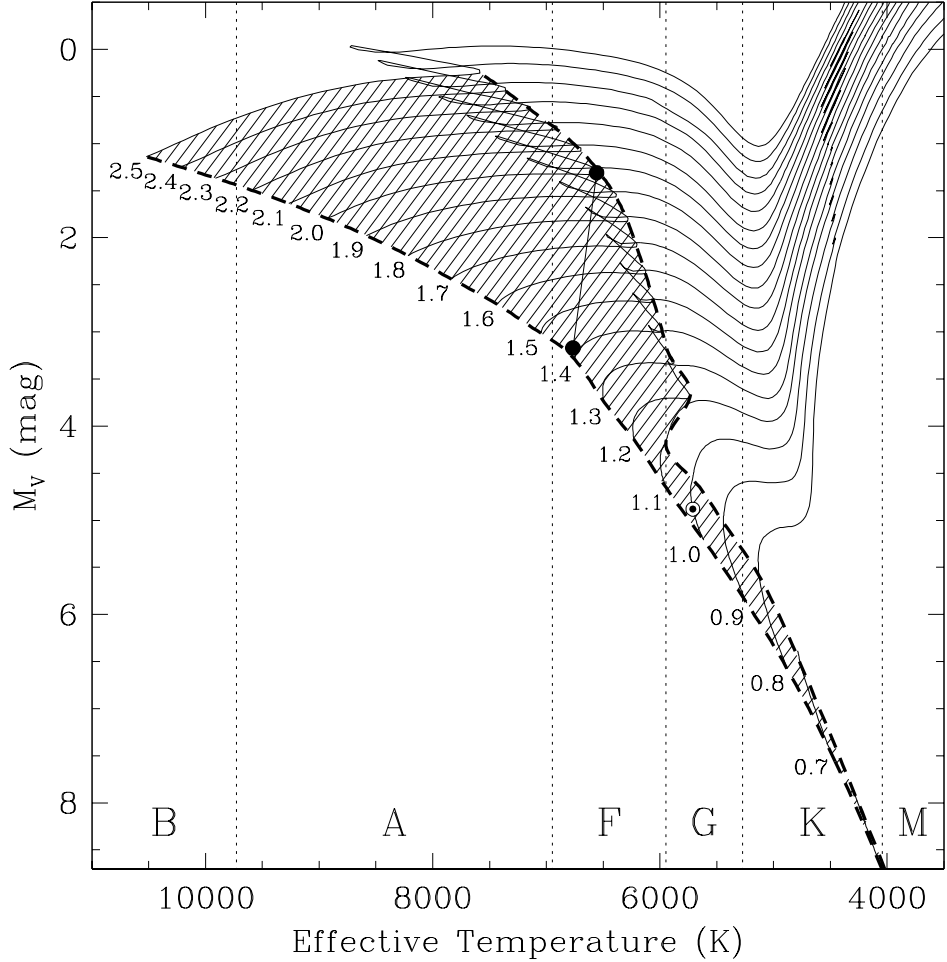}
\vskip -0.0in
 \figcaption[Torres.fig10.ps]{Evolutionary tracks from the series of
models by \cite{Yi:03} for solar metallicity, with the mass of each
track labeled on the left in solar masses. Thick dashed lines indicate
the ZAMS (left) and the red edge of the main-sequence phase (right).
The shaded area in between is where evolution makes stars
intrinsically brighter (by up to 2 magnitudes or more) while still on
the main sequence, and leaves ample room for a fainter unevolved star
to go undetected in the observations, producing a blend. We indicate
the location of the Sun on the 1.0~M$_{\sun}$ track for reference, as
well as the positions of star~1 and star~3 in OGLE-TR-33 (from
Fig.~\ref{fig:hr}) connected by a line. Approximate spectral type
boundaries are indicated along the bottom.\label{fig:hideblends}}
 \end{figure}

The situation is depicted more clearly in Figure~\ref{fig:hideblends},
where we show evolutionary tracks for solar metallicity in a diagram
of absolute visual magnitude as a function of effective
temperature\footnote{We have chosen for this illustration the models
by \cite{Yi:03} over those of \cite{Girardi:00}, used earlier, merely
for their higher resolution near the turnoff.}.  The heavy dashed
lines indicate the Zero Age Main Sequence (ZAMS) on the left, and the
red edge of the main sequence on the right, where stars have nearly
exhausted their hydrogen fuel and are about to undergo the brief
collapse phase that gives rise to the ``blue hook" feature. The shaded
area in between represents the full width of the main sequence band.
For most stars of spectral type G and later (spectral type boundaries
are indicated along the bottom of the figure) the vertical extent of
this region is only half a magnitude or less.  For hotter stars
beginning at spectral type F, however, there is a significant widening
of the main-sequence such that the vertical extent can exceed 2
magnitudes. It is precisely in this region of the H-R diagram that
OGLE-TR-33 is located. Because stars here can evolve to be much
brighter than unevolved stars of the same age, this leaves ample room
for a fainter eclipsing binary along the same line of sight to go
unnoticed.  To make matters worse, F stars are typically very common
in magnitude-limited samples such as OGLE-III, for the very same
reason that they are brighter. 

In blend cases in which star~1 and star~3 have very different
effective temperatures (and colors) photometric follow-up in multiple
passbands can be an efficient way to discover the false positive,
since the transit depth will depend on wavelength \citep[see,
e.g.,][]{Kotredes:04}. However, blends composed of stars of similar
temperature will \emph{not} show a significant color dependence, and
once again OGLE-TR-33 is one such example.  Based on the above we
predict that blends with no significant color dependence to the
eclipse depth may be more common than previously thought among F
stars. Therefore, special care is required when investigating F stars
with transit signatures, as the signs of the eclipsing binary may not
be immediately obvious either in the photometry or the spectroscopy. 
	
\section{Concluding remarks}
\label{sec:remarks}

Transit searches for extrasolar planets have attracted a great deal of
attention in recent years, driven by the excitement after the
discovery of HD~209458, and more recently OGLE-TR-56, OGLE-TR-113, and
OGLE-TR-132 \citep{Konacki:03a, Bouchy:04, Konacki:04}. Since no other
examples as bright as HD~209458 ($V = 7.65$) have been found to date,
focus has quickly shifted to denser fields and fainter stars.  As
researchers involved in these searches have learned by now, candidates
are relatively easy to find, but the large number of astrophysical
false positives is a very serious issue particularly for these fainter
targets.  Blending is perhaps the most insidious of these effects, and
is likely to be exacerbated in F stars because of the widening of the
main sequence.  The problem of finding true transiting planets around
faint stars has essentially become one of ruling out that a given
candidate can be anything else (``discovery by elimination").  Similar
or perhaps more challenging difficulties will be faced by future space
projects designed to search for even smaller planets, such as NASA's
{\it Kepler\/} mission. 

In the preceding sections we have shown that the complete set of
observations available for the transiting planet candidate OGLE-TR-33
can, in fact, be adequately explained as being the result of a fairly
typical blend scenario, and therefore that the photometric signature
that originally drew attention to it is very unlikely to be due to a
planet crossing in front of the star.  In this case the configuration
is a hierarchical triple system with an age of $\sim$1~Gyr composed of
an eclipsing binary (\ion{F4}{5} + \ion{K7-M0}{5}) and a slightly
evolved and brighter F6 star that dilutes the light of the close pair.
The self-consistent model we have constructed is able to account for
all photometric and spectroscopic constraints, including the detailed
morphology of the light curve (depth, duration, and shape), as well as
the composite nature of the spectrum, the velocity amplitude of
star~1, and its brightness relative to the main star. 

Achieving this level of agreement is of course not normally the goal
of follow-up observations of transit candidates. Rather, our
motivation here has been to show how a detailed modeling of the light
curve can provide important clues as to what other effects to look for
in the data that might produce compelling evidence against the planet
hypothesis. 

The scenario found to reproduce the observations for this particular
candidate is but one in an enormous variety of possible configurations
involving an eclipsing binary and another star. Other such examples
have been reported recently, e.g., by \cite{Charbonneau:04} from
another search program.  OGLE-TR-33 is especially interesting in that
it serves to highlight how subtle the signs of a blend can be ---in
this case the relatively small variations in the profiles of the
spectral lines. Had these variations been smaller (or our
spectroscopic observations of lower quality) they might have been
overlooked. The importance of careful follow-up cannot be overstated,
and alternate explanations should always be examined. If a candidate
with transit-like events of the right characteristics (shape, depth,
duration) shows no other evidence of contamination from an eclipsing
binary (spectroscopic or otherwise), can it \emph{still} be a blend?
This possibility must be thoroughly explored before a claim can be
made with any confidence.  Qualitative arguments or
``back-of-the-envelope" calculations are generally insufficient, and
we believe that analytical tools such as the one we have developed and
applied in this paper can be extremely helpful to guide the
investigation.  In a forthcoming paper we will explore quantitatively
the effects of blend scenarios on spectral line asymmetries and radial
velocities measured by cross-correlation. 

\acknowledgements

We are grateful to A.\ Udalski and the OGLE team for numerous
contributions to this project, and to the referee for helpful
comments.  Some of the data presented herein were obtained at the W.\
M.\ Keck Observatory, which is operated as a scientific partnership
among the California Institute of Technology, the University of
California and the National Aeronautics and Space Administration. The
Observatory was made possible by the generous financial support of the
W.\ M.\ Keck Foundation.  G.T.\ acknowledges support for this work
from NASA's {\it Kepler\/} mission, STScI program GO-9805.02-A, the
Keck PI Data Analysis Fund (JPL 1257943), and NASA Origins grant
NNG04LG89G.  M.K.\ gratefully acknowledges the support of NASA through
the Michelson fellowship program.  S.J.\ thanks the Miller Institute
for Basic Research in Science at UC Berkeley for support through a
research fellowship. This research has made use of NASA's Astrophysics
Data System Abstract Service.

\clearpage

\begin{deluxetable}{cccc}
\tablenum{1}
\tablewidth{18pc}
\tablecaption{Radial velocities measurements for star~1 and star~3 in
OGLE-TR-33\tablenotemark{a}.\label{tab:rvs}}
\tablehead{
\colhead{HJD} & \colhead{RV star~1} &
\colhead{RV star~3} \\
\colhead{(2,400,000+)} & \colhead{($\kms$)} &
\colhead{($\kms$)} }
\startdata
 52480.8503 &  $-$89.12 &  $-$28.16 \\
 52481.9398 &  $+$32.92 &  $-$29.45 \\
 52482.9403 &  $-$91.24 &  $-$27.95 \\
 52483.9357 &  $+$32.78 &  $-$29.65 \\ 
\enddata 
\tablenotetext{a}{Velocities in the barycentric frame. Estimated
errors are 1~\kms\ and 0.5~\kms\ for star~1 and star~3, respectively,
except for the first date (weak exposure), in which the uncertainties
are twice as large.} 
\end{deluxetable}

\clearpage
 
\clearpage
 
\clearpage
 
\clearpage
 
\clearpage
 
\clearpage
 
\clearpage
 
\clearpage
 
\clearpage
 
\clearpage
 
\end{document}